\documentclass[superscriptaddress,twocolumn]{revtex4}
\usepackage{amsmath,lscape,epsfig}
\begin{document}
\title{The importance of the mixed phase in hybrid stars built with the 
Nambu-Jona-Lasinio model}
\author{M.G. Paoli}                     
\affiliation{Depto de F\'{\i}sica - CFM - Universidade Federal de Santa
Catarina - Florian\'opolis - SC - CP. 476 - CEP 88.040 - 900 - Brazil}
\author{D.P. Menezes}                     
\affiliation{Depto de F\'{\i}sica - CFM - Universidade Federal de Santa
Catarina - Florian\'opolis - SC - CP. 476 - CEP 88.040 - 900 - Brazil}
\begin{abstract}
We investigate the structure of hybrid stars based on two different 
constructions: one is based on the Gibbs condition for phase coexistence and
considers the existence of a mixed phase (MP), and the other is based on the 
Maxwell construction and no mixed phase is obtained.
The hadron phase is described by the non-linear Walecka model (NLW) 
and the quark phase by the Nambu-Jona-Lasinio model (NJL).
We conclude that the masses and radii obtained are model dependent
but not significantly different for both constructions.%
\end{abstract}

\maketitle

\vspace{0.50cm}
PACS number(s): {24.10.Jv, 26.60+c}
\vspace{0.50cm}

\section{Introduction}
\label{intro}

Understanding the processes involved in the supernova explosions, in the
creation of stellar compact objects and in their temporal evolution requires
a huge multidisciplinary effort with investigations in areas as distinct as
nuclear and particle physics, thermodynamics, quantum field theory and 
astrophysics. 

In the present work we concentrate on the description of neutron stars. 
From very low densities up to the high densities present in their core, the 
constitution of these compact objects is a great source of speculation. 
At low densities there can be neutrons, protons, electrons and possibly 
neutrinos (at finite temperatures). At high densities, 
stellar matter can be much more complex, including hyperons, kaons and even
deconfined quarks.

Many works considering the construction of equations of state (EoS) used to
describe compact objects have already been done \cite{prakash,glen}. 
Once a certain EoS is obtained, it serves as input to the
Tolman-Oppenheimer-Volkoff equations (TOV) \cite{tov} and the output gives
the structure of the compact stars, characterized by their mass and radius.
An appropriate EoS or an inadequate one can only be chosen or ruled out once
astronomical observations are used as constraints.
Although some observational results are known, many uncertainties exist.
It is still unknown whether the neutron stars are composed only of hadrons and 
leptons, necessary to ensure chemical equilibrium and charge neutrality 
\cite{hadronicas}, if they are quark stars \cite{quarkionicas} or even
hybrid stars, containing both hadron and quark matter in their interior
\cite{schaff,lugones,outros,hibridas}. 
Each one of these possibilities is represented by a great 
variety of relativistic and even non-relativistic models used to built the EoS.

We next investigate hybrid stars only, whose existence is a source of intense
discussions in the literature \cite{glen,schaff,lugones,outros,hibridas}.
The discussion presented in \cite{lugones} is particularly interesting because
the existence of quark stars is shown to be questionable within the
calculations performed (which depend strongly on a specific parametrization).
Moreover, it is also pointed out that the possibility of a mixed population 
(or hybrid stars) is compatible with the calculations 
of model dependent quark matter nucleation, what reinforces the interest
in the calculations of hybrid stars as compact objects. Recent calculations
show the importance of the nucleation mechanism in the process of phase 
transition from hadronic to quark matter \cite{constanca,fraga}.

The main reason for the present work is the fact that many astrophysicists
claim that the mixed phase is only a hypothetical choice and cannot be checked.
Moreover, some authors calculated macroscopic quantities as radii and
masses for hybrid stars with and without the mixed phase and concluded that the
differences were not significant \cite{vos,maruyama} or that the region 
corresponding to the hadron-quark mixed phase is too narrow \cite{vos2}.
Although hybrid stars have been obtained with different combinations of models
for the hadron and the quark phases, most of the discussions on the use 
of Gibbs and Maxwell constructions have been based on the MIT bag model 
\cite{bag} for the description of the quark phase. 
The MIT bag model \cite{bag} is a very simple model that does not 
reproduce some of the necessary features of QCD at high densities, as chiral 
symmetry, for instance. As it is easily checked on the literature, all 
results for compact stars are model dependent. Hence, before completely 
ruling out the need for the Gibbs construction and the consequent existence
of the mixed phase in hybrid stars, it is desirable that another 
calculation with a different model for the quark phase is considered.
That is the basis of the calculations and discussion that follows.

In the present paper, the hadron phase is described by the
non-linear Walecka model (NLW) \cite{sw} and the quark phase by the
Nambu-Jona-Lasinio  model (NJL) \cite{njl}. Two different constructions are 
made: one with a mixed phase (MP) and another without the mixed phase, where
the hadron and quark phases are in direct contact. In the first case, 
neutron and electron chemical potentials are continuous throughout the stellar
matter, based on the standard thermodynamical rules for phase coexistence known
as Gibbs conditions. In the second case, the electron chemical potential 
suffers a discontinuity because only the neutron chemical potential is imposed
to be continuous. The condition underlying the fact that only a single chemical
potential is common to both phases is known as Maxwell construction. 
In our approach we ignore surface and Coulomb effects for the structure in 
the mixed phase so the leptons are taken into account as free Fermi gases.
However, it is worthy pointing out that the energy density in mixed
phases should depend on the eletromagnetic and surface contributions and this is
commonly known as finite size effects. In \cite{maruyama,vos2,yatsutake} it
was shown that for a weak surface tension the EoS resembles the one obtained
with a Gibbs construction while for a strong surface tension, the Maxwell
construction was reproduced. Unfortunatelly, the surface energy coefficient
is not well described yet \cite{pasta_alpha}.
The differences between  stellar structures obtained with both constructions
are discussed through out the present paper. A similar calculation was done 
in \cite{yatsutake}, where 
the effects of different lepton fractions on protoneutron stars with trapped 
neutrinos were investigated. Although the result for zero temperature was also
presented, its validity when trapped neutrinos are enforced is only academic 
because the neutrino mean free path at T=0 is larger than the neutron star 
radius. While in \cite{yatsutake} no hyperons were included in the 
hadronic phase, they are also taken into account in the present paper 
for two parametrizations of the NLW model. Notice, however, that $s$ quarks
were also considered in the quark phase described in \cite{yatsutake}.

In works involving quark \cite{quarkionicas} or hybrid stars \cite{hibridas}, it
is seen that the NJL model gives results that are quite different from the ones
obtained with the MIT model. The fact that the NJL model incorporates chiral 
symmetry and that the strange quark appears only in densities much higher than
the $u$ and $d$ quarks are the main reasons for the differences. Hence, the
calculations for the hybrid stars are here done with the NJL model so the
previous conclusions on the mixed phase are confirmed or refuted.
The consequences of the inclusion of the $s$-quark in the NJL model at quite
high densities is also seen once a comparison between the two versions of the
NJL model, i.e., SU(2) and SU(3) is performed. Whenever the SU(2) version of 
the NJL is used to describe quark matter, the corresponding hadron phase is 
strangeness free, i.e., no hyperons are considered. Two parameter sets are used
for each case considered so that the model dependence can be established.

The paper is organized as follows: In Sec. 2 we show the lagrangian densities 
of the models considered and describe the formalism used; in Sec. 3 we 
present and discuss the results; in Sec. 4 we draw our final conclusions.
\section{The Formalism}
\label{form}

We next give some of the main equations related to the two models used in 
our investigation. Detailed calculations are extensively available in the 
literature and hence are omitted in the present paper. Two possible systems 
are studied: one 
comprehends 8 baryons in the hadron phase and 3 quarks in the quark phase and
the other includes only protons and neutrons in the hadron phase and the 
corresponding $u$ an $d$ quarks in the quark phase. In most cases, our studies
refer to a hadron matter with all 8 baryons and a quark matter with
the 3 possible quarks. 
\subsection{The NJL model}
\label{njl}

The NJL model is defined by the lagrangian density
$$ {\cal L}_{NJL}=\bar{q}(i\gamma^\mu \partial_\mu - m)q + g_S \sum_{a=0}^{8}
[(\bar{q}\lambda^a q)^2 + (\bar{q}i\gamma_5 \lambda^a q)^2]$$
\begin{equation}
- g_D \{ det[\bar{q}_i(1+\gamma_5)q_j] + det[\bar{q}_i(1-\gamma_5)q_j] \},
\label{lagnjl}
\end{equation}
where $q = (u,d,s)$ are the quark fields and $\lambda_a$ $(0 \leq a \leq 8)$, 
are the U(3) flavour matrices. The model parameters are the current quark mass 
matrix $m = diag ( m_u , m_d , m_s )$, the coupling constants $g_S$ and $g_D$, 
and the cutoff in 3-momentum space $\Lambda$. 

The thermodynamical potential density is given by 
$\Omega=\varepsilon - \sum_{i} \mu_i \rho_i - \Omega_0,$ 
where the energy density is
$$ \varepsilon = -2N_c\sum_{i}\int \frac{d^3 p}{(2\pi)^3}\frac{p^2 + m_i M_i}
{E_i}\theta(\Lambda^2-p^2)$$
\begin{equation}
-2 g_S \sum_{i=u,d,s}\left \langle \bar{q_i}q_i \right \rangle^2 + 2 g_D \left \langle \bar{u}u \right \rangle\left \langle \bar{d}d \right \rangle\left \langle \bar{s}s \right \rangle-\varepsilon_0.
\label{enernjl}
\end{equation}
In the above expressions, $N_c=3$, $E_i=\sqrt{p^2+M_i^2}$, $\mu_i(\rho_i)$ is 
the chemical potential 
(number density) of particles of type $i$, and $\varepsilon_0$ and 
$\Omega_0$ are included in order to ensure $\varepsilon=\Omega=0$ in the 
vacuum. The quark condensates and the quark densities are defined, for each 
of the flavors $i=u,d,s$,  respectively, as:
\begin{equation}
\langle\bar q_i\, q_i\rangle = -2 N_c\, \int {d^3 p\over (2\pi)^3}
{M_i\over E_i} \theta (\Lambda^2 -p^2), \label{6}
\end{equation}
\begin{equation}
\rho_i\,=\, \langle{q_i}^{\dagger}\, q_i\rangle = \frac{{P_F}_i^3}{3 \pi^2}.
\label{7}
\end{equation}

Minimizing the thermodynamical potential $\Omega$ with respect to 
the constituent quark masses $M_i$ leads to three gap equations for 
the masses $M_i$
\begin{equation}
M_i\,=\,m_i\,-4\,g_S\,\langle\bar q_i\, q_i\rangle\,+\,2\,g_D\,\langle\bar
q_j\, q_j\rangle\langle\bar q_k\, q_k\rangle\,,\label{8}
\end{equation}
with cyclic permutations of  $i,\, j,\, k$.

The pressure can be found from
\begin{equation}
P = -\Omega = -\varepsilon + \sum_{i} \mu_i \rho_i + \Omega_0.
\end{equation}
The relations between the chemical potentials of the different particles 
required by $\beta$- equilibrium are given by 
\begin{equation}
\mu_s = \mu_d = \mu_u + \mu_e,\ \ \mu_e=\mu_\mu
\label{betaquark}
\end{equation}
and for charge neutrality we must impose
\begin{equation}
\rho_e+\rho_\mu=\frac{1}{3}(2\rho_u-\rho_d-\rho_s).
\label{chargequark}
\end{equation}

In order to obtain the NJL SU(2) model we just need to neglect the terms 
related to the strange quark in equations 2, 4 and 5.
It means that $\langle \bar{s}s \rangle = \mu_s = \rho_s = 0$.

The parameter sets of the NJL model used in the present work are given in 
Table 1.

\subsection{The non-linear Walecka model}
\label{nlw}

The lagrangian density for the NLW model reads
$${\cal L}_{NLW}=\sum_B \bar{\psi}_B \left[ \gamma_\mu \left( i\partial^{\mu}
-g_{vB} V^{\mu} - g_{\rho B} {\vec \tau} \cdot {\mathbf b}^\mu 
\right)\right.$$
$$\left.-(M_B-g_{s B} \phi)\right]\psi_B 
+\frac{1}{2}(\partial_{\mu}\phi\partial^{\mu}\phi
-m_s^2 \phi^2)
-\frac{1}{3!}\kappa \phi ^{3}
-\frac{1}{4!}\lambda \phi ^{4}$$
\begin{equation}
-\frac{1}{4} \Omega_{\mu\nu}\Omega^{\mu\nu}
+\frac{1}{2} m_v^2 V_{\mu}V^{\mu}
-\frac{1}{4}\mathbf B_{\mu\nu}\cdot\mathbf B^{\mu\nu}
+\frac{1}{2} m_\rho^2 \mathbf b_{\mu}\cdot \mathbf b^{\mu},
\label{lagwalecka}
\end{equation}
with $\Omega_{\mu \nu} = \partial_\mu V_\nu - \partial_\nu V_\mu$
and $\mathbf{B}_{\mu \nu} = \partial \mu \mathbf{b}_\nu - \partial_\nu 
\mathbf{b}_\mu - g_\rho(\mathbf{b}_\mu \times \mathbf{b}_\nu).$ 
The hyperon coupling constants are defined as 
$x_i = \frac{g_{iB}}{g_{i}}, \quad i=s,v,\rho$.

In a mean field approximation the energy density reads
$$\varepsilon = \frac{\gamma}{2\pi^2} 
\sum_{B}\int_{0}^{K_{FB}}p^2dp \sqrt{p^2+M_B^{*2}}$$
\begin{equation}
+\frac{m_v^2}{2}V_0^2
+\frac{m_\rho^2}{2}b_0^2
+\frac{m_s^2}{2}\phi_0^2
+\frac{\kappa}{6}\phi_0^3
+\frac{\lambda}{24}\phi_0^4,
\end{equation}
and the pressure becomes
$$P = \frac{\gamma}{6\pi^2} \sum_{B}\int_{0}^{K_{FB}}
\frac{p^4dp}{\sqrt{p^2+M_B^{*2}}}$$
\begin{equation}
+\frac{m_v^2}{2}V_0^2
+\frac{m_\rho^2}{2}b_0^2
-\frac{m_s^2}{2}\phi_0^2
-\frac{\kappa}{6}\phi_0^3
-\frac{\lambda}{24}\phi_0^4,
\end{equation}
where $\gamma=2$ is the spin degeneracy factor.

The conditions of chemical equilibrium are also imposed through the two 
independent chemical potentials $\mu_n$ and $\mu_e$ and it implies that
$$\mu_{\Sigma^0}=\mu_{\Xi^0}=\mu_{\Lambda}=\mu_n,$$
$$\mu_{\Sigma^-}=\mu_{\Xi^-}=\mu_n + \mu_e, \\$$
\begin{equation}
\mu_{\Sigma^+}=\mu_p=\mu_n - \mu_e.
\label{betahadron}
\end{equation}
For the charge neutrality, we must have
\begin{equation}
\sum_{B}q_B \rho_B + \sum_{l} q_l \rho_l = 0,
\label{chargehadron}
\end{equation} 
where $q_B$ and $q_l$ stand, respectively, for the electric charges 
of baryons and leptons.

When the system is constituted only of protons and neutrons, the sum on the 
baryons appearing in the equations above are restricted to the nucleons and the
only condition for chemical equilibrium is the last one in (\ref{betahadron}).

Two sets of parameters were chosen and they are given in Table 2. 
The following parameters are equal for both sets: 
$x_s=0.7$, $x_v=x_\rho=0.783$, $m_s=400$ MeV, $m_v=783$ MeV
and $m_\rho=770$ MeV.

%
\subsubsection{Comments on the inclusion of the leptons}
\label{leptons}

As we are dealing with neutral stellar matter in $\beta$- equilibrium in both 
quark and hadron phases according to eqs. 
(\ref{betaquark})-(\ref{chargequark}) and 
(\ref{betahadron})-(\ref{chargehadron}) respectively, 
the electrons and muons have to be introduced. They are normally included as 
free Fermi gases obeying the following lagrangian density:
\begin{equation}
{\cal L}_l=
\sum_{l} \bar{\psi}_l(i\gamma_\mu \partial^\mu- m_l)\psi_l,  \quad
l=e^-,\mu^-.
\end{equation}
Expressions for energy density and pressure in a MFT become:
\begin{equation}
\varepsilon =
\\ \frac{1}{\pi^2}\sum_{l}\int_{0}^{K_{Fl}}p^2 dp \sqrt{p^2+m_l^2},
\end{equation}
and
\begin{equation}
P= \frac{1}{3 \pi^2}\sum_{l}\int_{0}^{K_{Fl}}\frac{p^4dp}{\sqrt{p^2+m_l^2}}.
\end{equation}
\subsection{Hybrid stars with mixed phase}
\label{with}

We next build a mixed phase (MP) constituted of hadrons and quarks, which 
interpolates between the hadron (HP) and the quark phase (QP).
In the mixed phase charge neutrality is not imposed locally but only 
globally. This means that quark and hadron phases are not neutral 
separately, but rather, the system prefers to rearrange itself so that
$$\chi \rho_c^{QP}+(1-\chi)\rho_c^{HP}+\rho_c^l=0,$$
where $\rho_c^{iP}$ is the charge density of the phase $i$, $\chi$ is the volume fraction occupied by the quark phase,
and $\rho_c^l$ is the electric charge density of leptons. According to the Gibbs conditions for phase coexistence, the 
neutron chemical potentials, the electron chemical potentials and pressures have to be identical in both phases,
i.e. \cite{glen},
$$\mu^{HP}_n=\mu^{QP}_n,\quad \mu^{HP}_e=\mu^{QP}_e \quad {\rm and} \quad P^{HP}=P^{QP}.$$
As a consequence, the energy density and total baryon density 
(no leptons included) in the mixed phase read
\begin{equation}
\langle \varepsilon \rangle = \chi \varepsilon^{QP}+(1-\chi)\varepsilon^{HP}+\varepsilon^l
\end{equation}
and
\begin{equation}
\langle \rho \rangle = \chi \rho^{QP}+(1-\chi)\rho^{HP}.
\end{equation}

\subsection{Hybrid stars without mixed phase}
\label{without}

Much simpler than the case above, we just need to find the point where
$$ \mu^{HP}_n=\mu^{QP}_n \quad {\rm and} \quad P^{HP}=P^{QP},$$
and then construct the EoS. In this case the electron chemical potential 
suffers a discontinuity when passing from the hadron to the quark phase
as expected from the simple use of the Maxwell conditions.
\section{Results}
\label{results}

In the graphs shown next SU(3) stands for the quark phase taking into
account the strange quark and SU(2) represents the NJL model without 
the strange quark. GM1 and GM3 represent the hadron 
phase with their respective set of parameters. Systems without strangeness
are described by protons and neutrons in the hadron phase and quarks
$u$ and $d$ in the quark phase. Systems with strangeness also accommodate the 
hyperons in the hadron phase and quark $s$ in the quark phase.

In Fig. 1 the EoS for the pure hadron and pure quark matter are shown. 
They are the base to built the EoS of the hybrid stars. 
The kick, or variation, in curvature in the quark system is due to the 
appearance of the $s$-quark.

In Figs. 2 and 3 the EoS for the two types of hybrid 
stars are shown with parametrizations GM1 and GM3 respectively for the hadron
phase. Two parametrizations (SU(3) HK and SU(3) RKH) are used for 
the quark phase in both figures. In both figures the hyperons and strange 
quarks are included.
The EoS of the hybrid stars with mixed phase are built by the superposition 
of the EoS for the quark and hadron matter, plus the EoS for the mixed phase.
The plateau in the EoS of the hybrid stars without mixed phase shows a vivid
phase transition from hadron to quark matter.
One can see that GM1 produces a much larger quark phase in both constructions 
while GM3 gives rise to a very large mixed phase when it is present and a quark 
phase much smaller than the hadron phase if a Maxwell construction is used.
This consideration is true independently of the parametrization used in the
quark phase, what means that the size of each phase is basically dependent on
the hadron phase parametrization, at least for the choices we have considered. 
This fact has obvious consequences in the constituents of the stellar matter.

In Figs. 4 and 5 we compare hybrid stars without the 
mixed phase built with the SU(2) and SU(3) NJL models, to understand the
role played by the strange quark.
Whenever  SU(2)set1 is used, the hadron phase is minimal and the EoS is 
practically given only by the quark phase. In all the other 
cases the size of each phase is mainly determined by the hadron phase 
parametrization, with small variations in the size of each phase. Notice that
the $u$ and $d$ quark vacuum masses are not identical, being the smallest for
the SU(2)set1 parametrization, as seen in Table 1.

We use all the EoS studied and commented before as input to the 
TOV equations to obtain the neutron star profiles that are shown in Fig. 
6, 7 and Table 3. The tails of the hadron and hybrid 
stars were obtained with
the insertion of the BPS EoS \cite{bps}. As expected, the maximum masses for
the hadron stars are larger than for the hybrid stars.
When the Maxwell construction is used, the
resulting mass-radius curve shows a kink, produced by the sharp transition in
the EoS. A similar result is shown in \cite{schramm} and it is easier to see in 
Fig. 7 , where we once more compare the hybrid stars without 
mixed phase built with the NJL SU(2) and SU(3) models. 
As a consequence of the EoS, GM1 and GM3 produce identical stellar profiles 
if only protons, neutrons and consequently quarks $u$ and $d$ are considered
with SU(2)set1. SU(2)set2 produces different results.
If only stars with strangeness are considered, GM1 always result in stars with
larger maximum masses and radii than GM3.

In Table 3 we also show the results for the central energy density
$\varepsilon_0$. $\varepsilon_{min}$ corresponds to the point where the hadron 
phase disappears, either because of the onset of a mixed phase (whenever 
$\varepsilon_{max}$ is also shown) or giving rise to the quark phase 
(otherwise). If strangeness are considered, in hybrid stars without 
mixed phase, 
the central energy density may lie within the hadron phase as in
GM3$\times$SU(3)HK and GM3$\times$SU(3)RKH. This means that if a phase 
transition to the quark phase occurs, the star becomes unstable.
On the other hand, if the parametrization GM1$\times$SU(3)RKH or 
GM1$\times$SU(3)HK is used, the central energy density lies in the quark phase.
If a mixed phase is considered, the central energy density 
always lies in its interior. As a consequence, the pure quark phase is never 
present. This is the only effect (not possible to infer from astronomical
observations) that depends strongly on the choice between Gibbs and Maxwell 
constructions. Nevertheless, it is well known that the strange quark condensate
is very large within the SU(3) NJL model and our results are a consequence of 
this behaviour. For this specific reason, we have also checked the results for
hybrid stars built without strangeness. 

If strangeness is not included, SU(2)set1 gives rise to hybrid stars with
central energy densities in the quark phase and SU(2)set2 shows unstable
solutions after the onset of the quark phase with the GM1 parametrization. 

Analyzing the results shown in Table 3 and based on the accuracy of our 
calculations and the experimental difficulties in the measurements of neutron 
stars radii, it is fair to say that the method used to built the EoS, i.e., 
the more rigorous Gibbs conditions or the simple use of the Maxwell 
construction give almost indistinguishable results for gravitational masses and
radii.

Finally, in Figs. 6 and 7 we have added three lines 
corresponding to observational constraints.
Some properties of the neutron stars are determinated by
measuring the gravitational redshift of spectral lines produced in neutron
star photosphere which provides a direct constraint on the mass-to-radius
ratio (M/R). A redshift of $z = 0.35$ from three different transitions of the
spectra of the X-ray binary EXO0748-676 was obtained in \cite{cottam1}. This
redshift corresponds to $M/R = 0.15~M_{\odot}/Km$. The top line corresponds 
to this constraint, whose validity remains controversial \cite{cottam2}. 
On the other hand, the 1E 1207.4-5209 neutron star, which is in the center of 
the supernova remnant PKS 1209-51/52 was also observed and two absorption 
features
in the source spectrum were detected \cite{sanwal1}. These features were
associated with atomic transitions of once-ionized helium in the neutron star
atmosphere with a strong magnetic field. This interpretation leads to a
redshift of the order of $z = 0.12 - 0.23$. This redshift imposes another 
constraint to the mass to radius ratio given by $M/R = 0.069~M_{\odot}/Km$ 
to $M/R = 0.115~M_{\odot}/Km$. This constraint is represented by the two lowest
lines.  One can see in Figs. \ref{fig:6} and \ref{fig:7} that all the curves 
obtained are consistent with the measurements of \cite{cottam1} and 
\cite{sanwal1} by crossing the 3 lines. 

\section{Conclusions}
\label{conclusions}

Assuming that hybrid stars are possible remnants of supernova explosions, their
constitution becomes important only if their macroscopic quantities can be 
constrained to astronomical observations. While some calculations practically 
exclude the existence of hybrid stars \cite{vos3} favoring quark stars, 
others tend to rule out quark stars and favor hybrid stars \cite{lugones}. 
All those conclusions are obviously model dependent and were reached based 
on the use of the MIT bag model to describe quark matter.

Hybrid stars have a hadron phase, in this paper described by the
non-linear Walecka model (NLW) \cite{sw} and a quark phase.
Instead of using the usual MIT bag model \cite{bag} to build the quark phase, 
we have opted to use the NJL model \cite{njl} to check some of the previous 
results on the existence of the mixed phase inside hybrid stars.

If the hadron phase is constituted of protons and neutrons, the 
corresponding quark phase has quarks $u$ and $d$ only and the SU(2) version 
of the NJL model is used. If the baryonic octet is
possible in the hadron phase, quarks $u$, $d$ and $s$ are present in the
quark phase described by the SU(3) NJL model.

We have concluded that the results are very model dependent, as expected.
The onset of a stable quark phase is practically ruled out. Our calculations
suggest that stable neutron stars are either of hadronic nature only or bear 
a mixed phase in their core.
Concerning the existence of the mixed phase (MP), one can see that the stellar 
measurable results calculated in the present paper (mass, radii, central 
energy density)
depend very little on the choice of the Maxwell or the Gibbs 
construction. Hence, it is reasonable to claim that the Maxwell construction
gives satisfactory results.

The effects of colour superconductivity are out of the scope of the present 
work, but it is important to mention that they may play an important role in 
the description of neutron star matter \cite{buballa}. The
colour-flavour-locked phase (CFL) could turn into a superconducting phase (2SC)
before matter is hadronized when we read the QCD phase diagram from high to low
densities (see a figures \cite{buballa,bursts}, for example).
This non-continuos transition from the CFL to the 2SC phase in the presence
of realistic strange quark masses would certainly affect the description
of hybrid stars.
\section*{ACKNOWLEDGMENTS}
This work was partially supported by CNPq (Brazil). The authors would like to
thank fruitful discussions with Dr. Constan\c ca Provid\^encia and two
anonymous referees for suggestions that improved the presentation of our 
results.

%


\newpage

\begin{figure}
\resizebox{8.5cm}{!}{%
\includegraphics{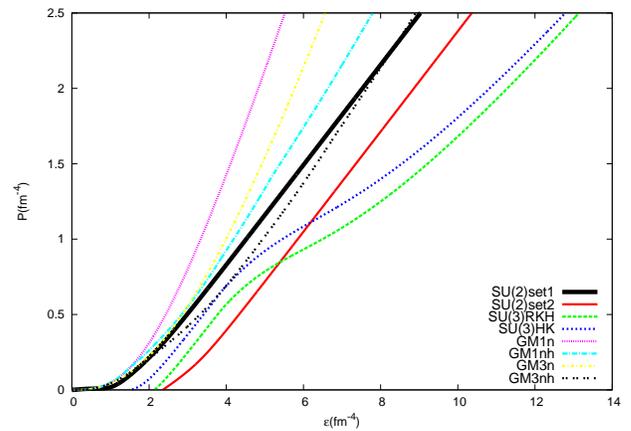}
}
\caption{EoS for the pure hadron and pure quark matter.}
\label{fig:1}
\end{figure}

%
%
%
%
%
%
%
%

\newpage
.
\newpage

\begin{center}
\begin{tabular}{cc}

\includegraphics[width=8.5cm]{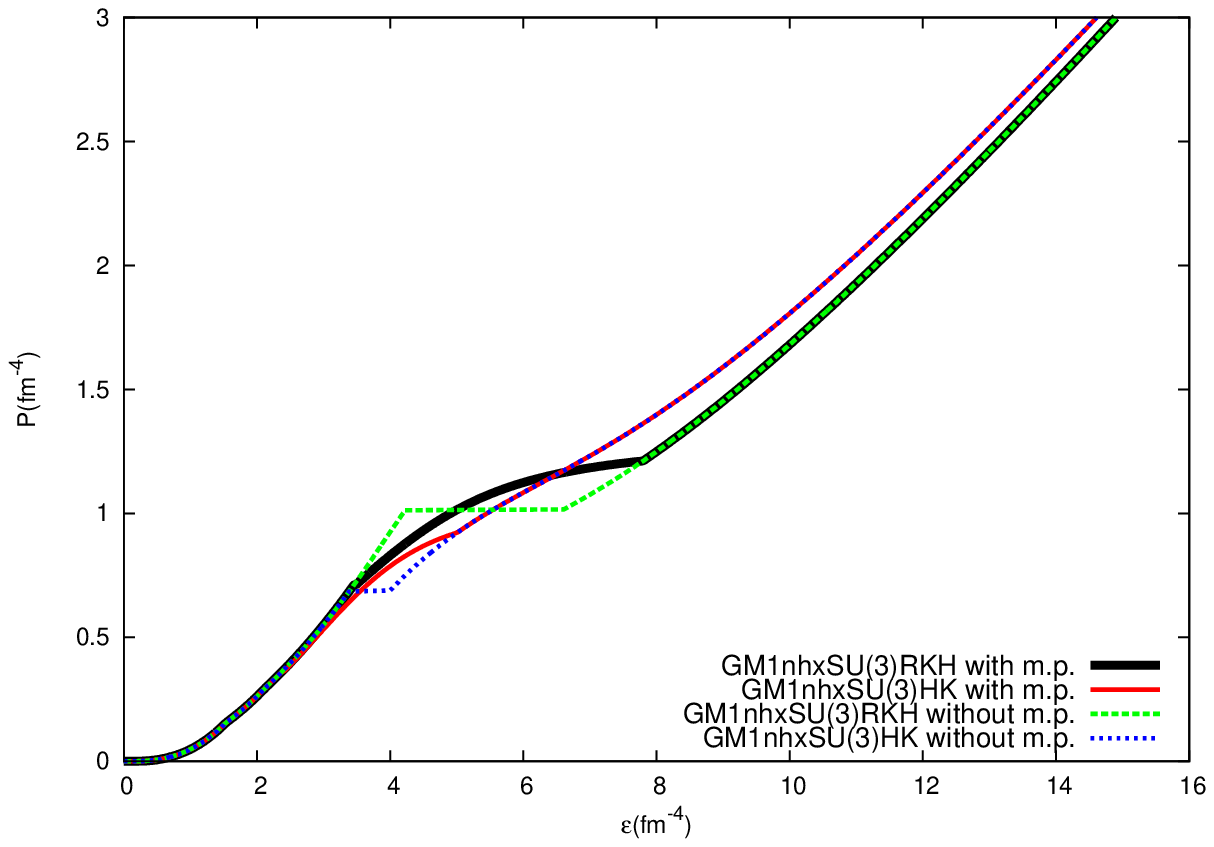}        & \includegraphics[width=8.5cm]{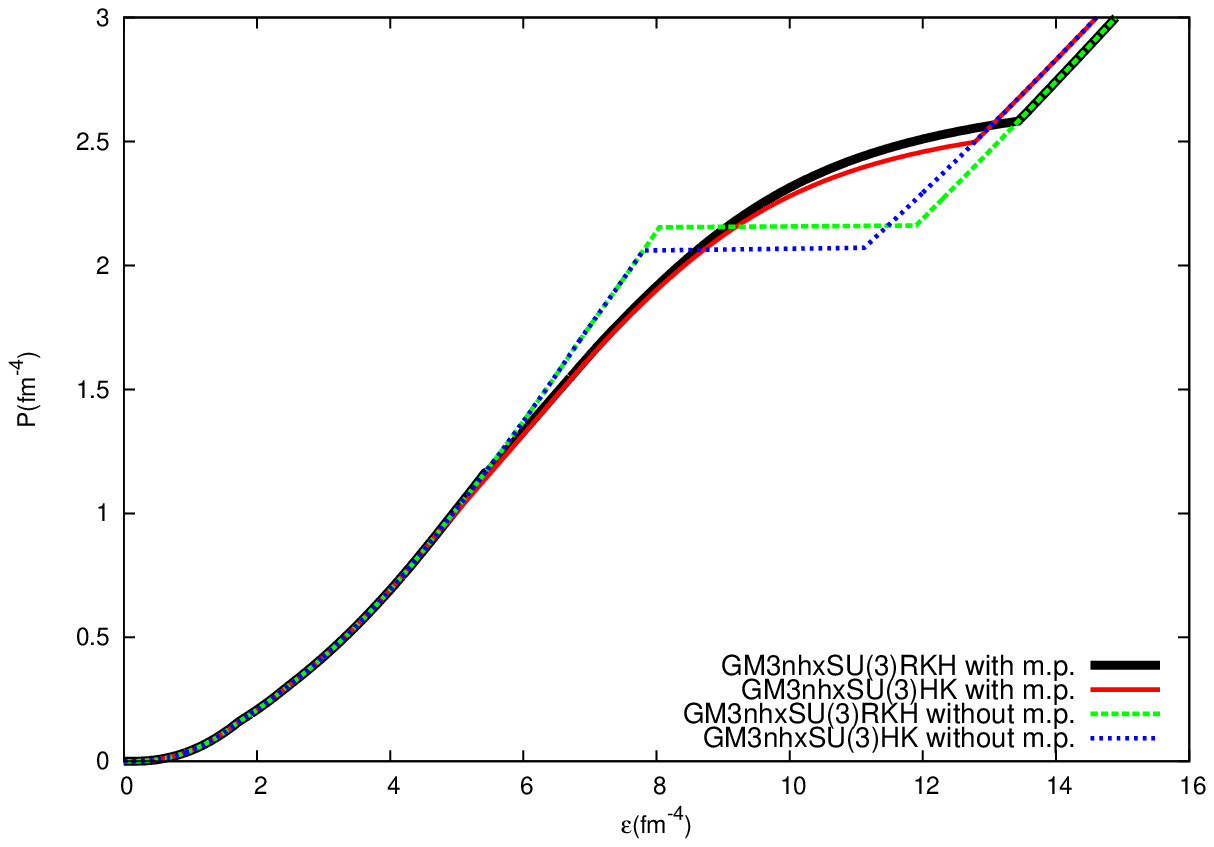}         \\
FIG. 2: EoS for the hybrid star with and without the mi-  & FIG. 3: EoS for the hybrid star with and without the mi-   \\
xed phase build with the GM1 parametrization, n stands  & xed phase build with the GM3 parametrization, n stands   \\
for nucleons only and nh for nucleons and hyperons.               &  for nucleons only and nh for nucleons and hyperons.                \\
\includegraphics[width=8.5cm]{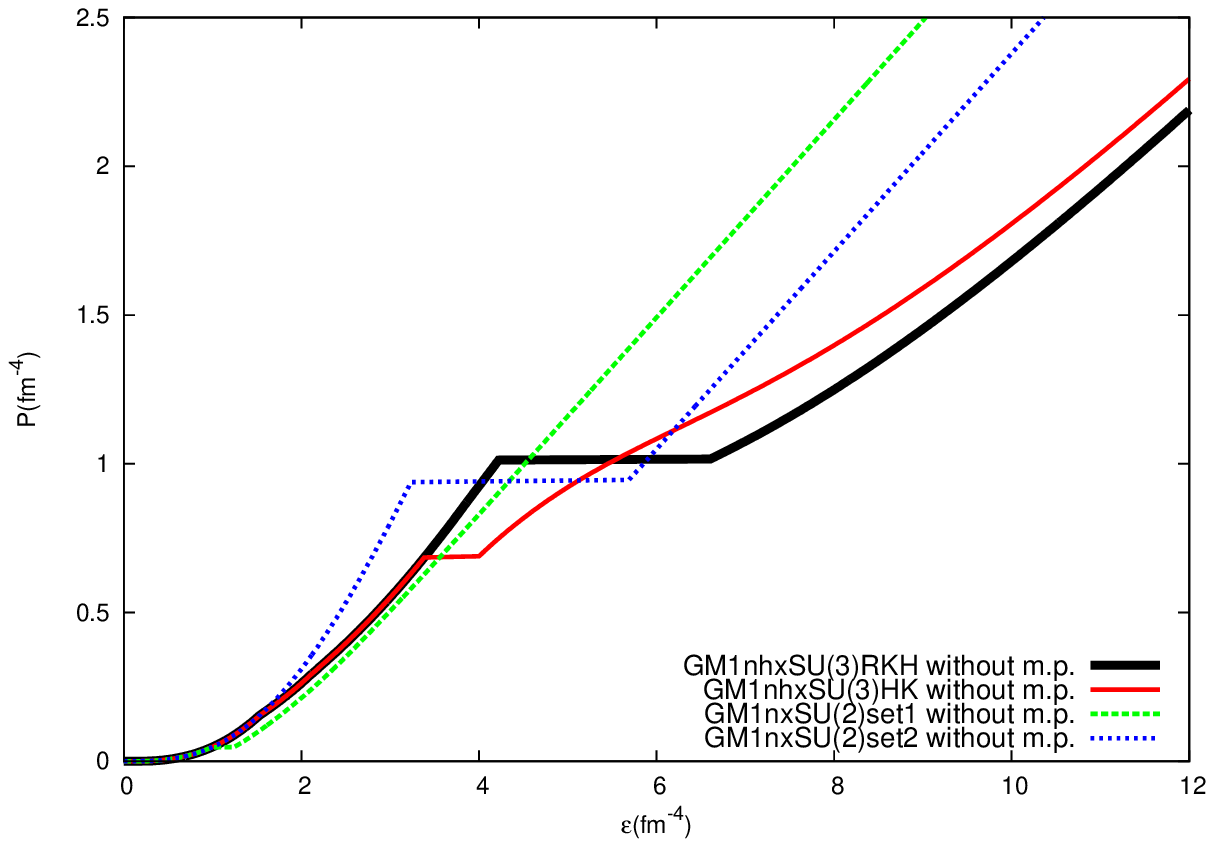}           & \includegraphics[width=8.5cm]{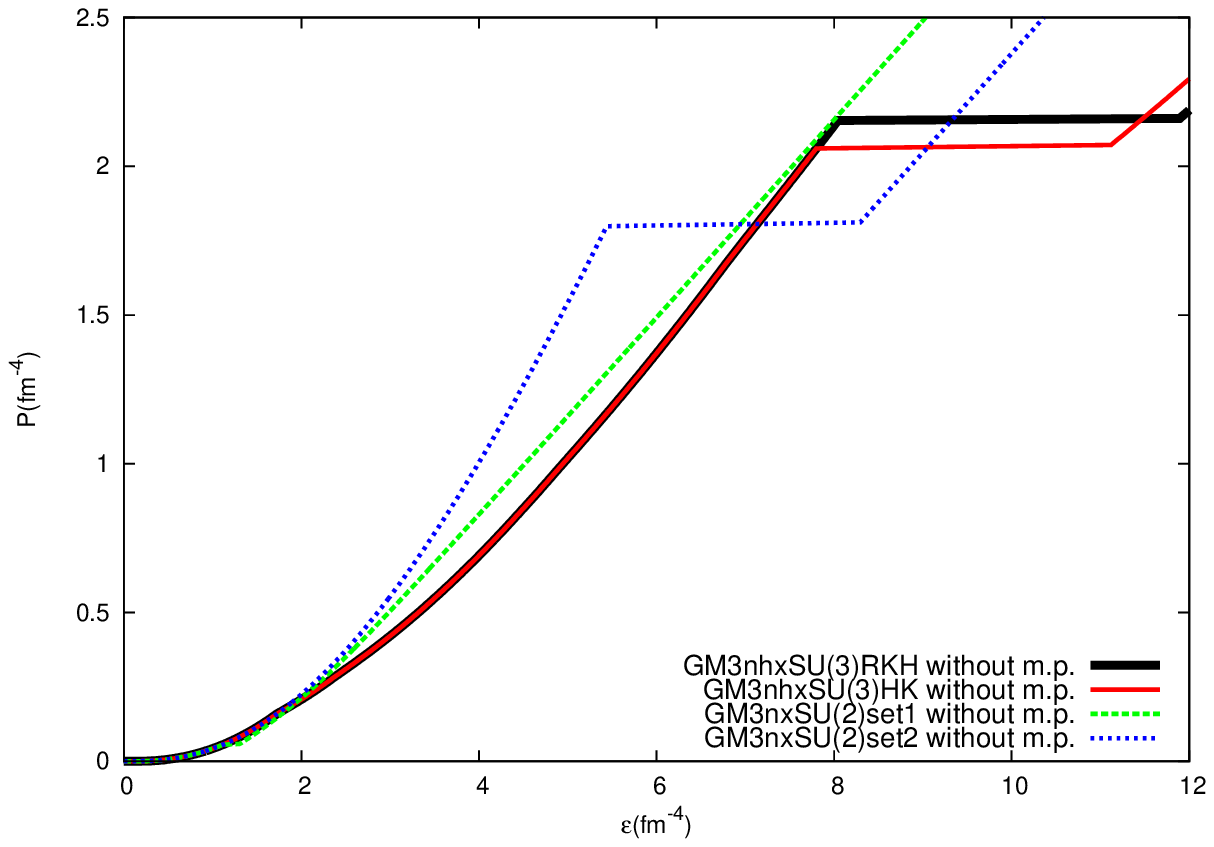}            \\
FIG. 4: EoS for the hybrid star without the mixed phase     & FIG. 5: EoS for the hybrid star without the mixed phase      \\
build with the NJL SU(3) and SU(2) model with the        & build with the NJL SU(3) and SU(2) model with the         \\
GM1 parametrization.                                            & GM3 parametrization.                                             \\ 
\includegraphics[width=8.5cm]{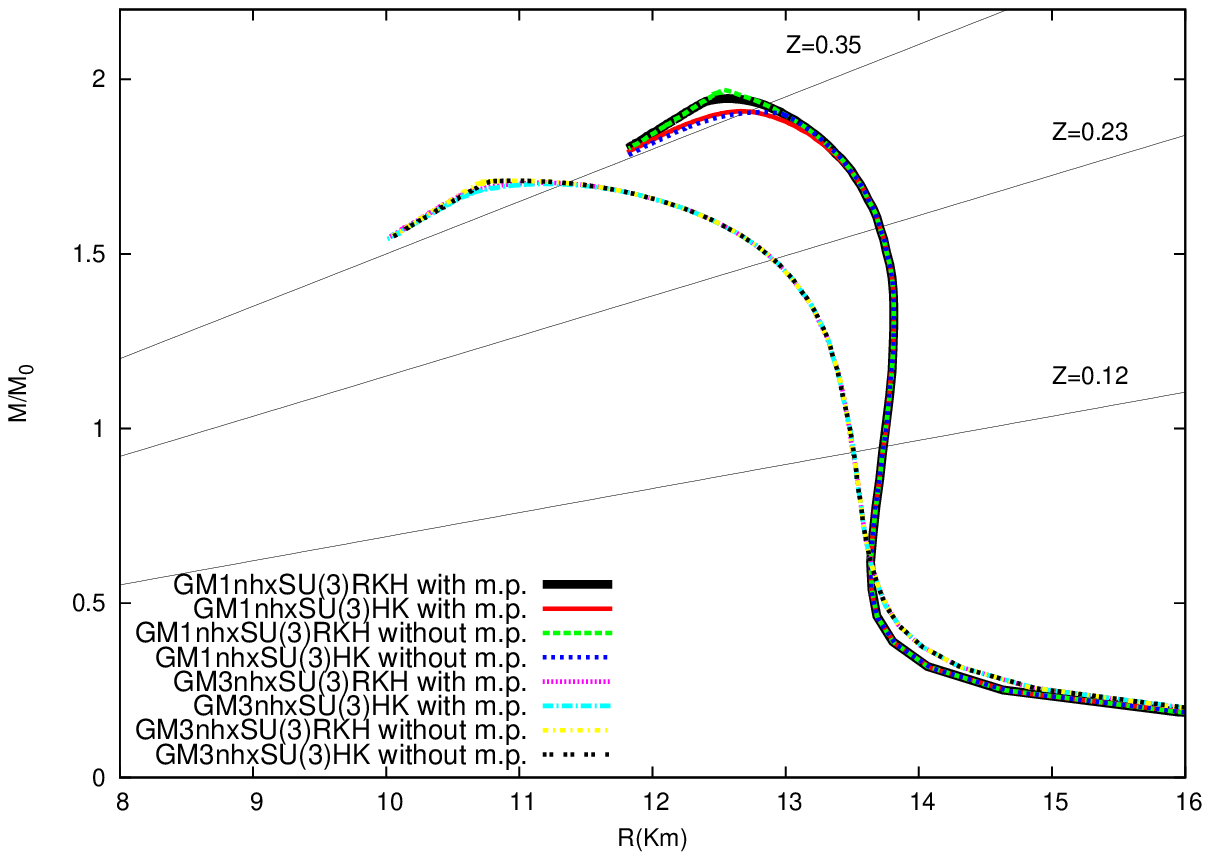}                 & \includegraphics[width=8.5cm]{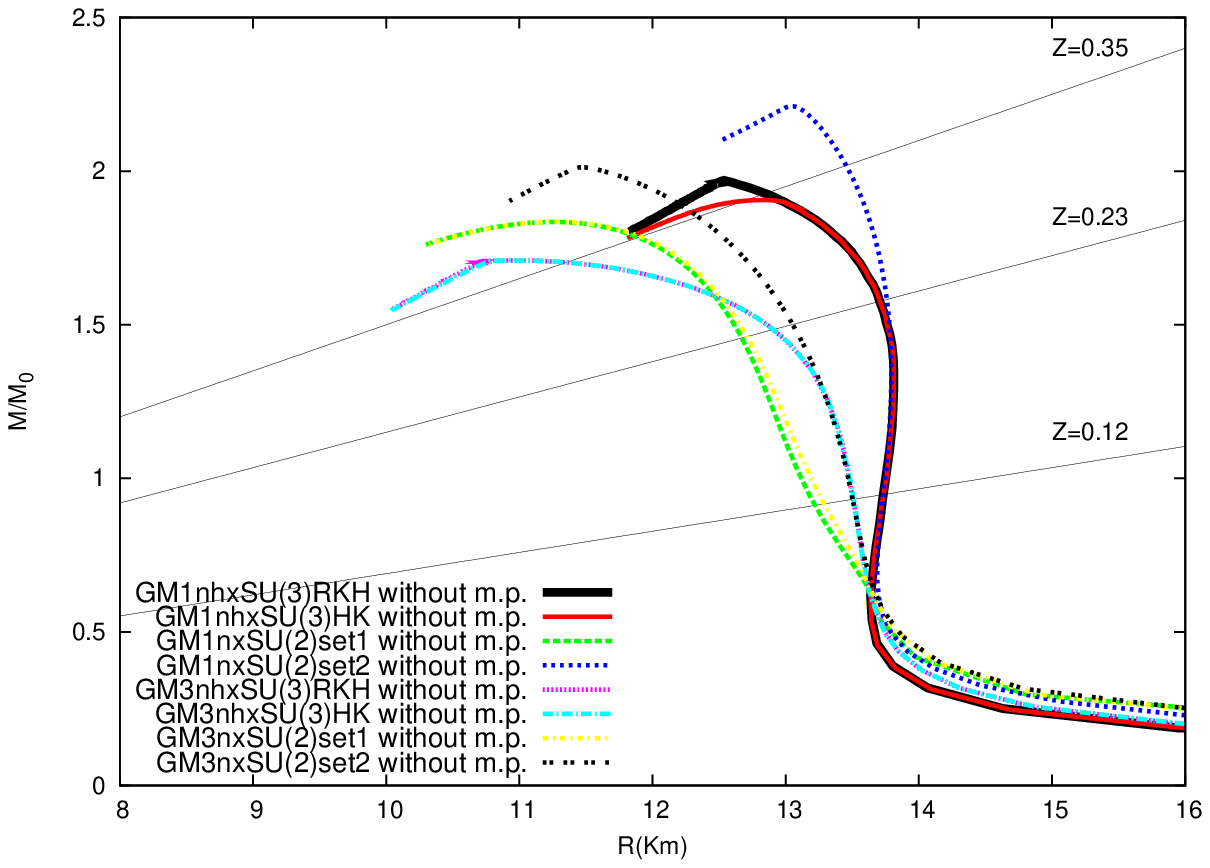}               \\
FIG. 6: Mass-radius curves for the hybrid stars with    & FIG. 7: Mass-radius curves for the hybrid stars without  \\
and without the mixed phase.                                    & the mixed phase build with the NJL SU(3) and SU(2)        \\
                                                        & model.
\end{tabular}
\end{center}


\begin{table}
\begin{center}
\begin{tabular}{lccccccc}
\hline
& $\Lambda$  &  &  & $m_{u,d}$ & $m_s$ & $M_{u,d}$ & $M_s$\\
Parameter set & (MeV)      &  $g_S\Lambda^2$ & $g_D\Lambda^5$ & (MeV) & (MeV) & (MeV) 
& (MeV)\\
\hline 
SU(2)set1\cite{buballa}& 664.3  & 2.06  & - & 5.0  & - & 300 & -    \\
SU(2)set2\cite{buballa}& 587.9  & 2.44  & - & 5.6  & - & 400 & -    \\
SU(3)HK\cite{buballa,HK}&631.4  & 1.835  & 9.29 & 5.5 & 135.7 & 335 & 527\\
SU(3)RKH\cite{buballa,RKH}& 602.3 & 1.835  & 12.36 & 5.5 & 140.7 & 367.7 & 
549.5\\
\hline
\end{tabular}
\end{center}
\caption{Parameter sets for the NJL SU(2) and SU(3) models.}
\end{table}

\begin{table}
\begin{center}
\begin{tabular}{lccccc}
\hline
                  & $(g_\sigma/m_s)^2$ & $(g_\omega/m_v)^2$ & $(g_\rho/m_\rho)^2$ &            &           \\
Parameter set             & $(fm^2)$      & $(fm^2)$      & $(fm^2)$            & $\kappa/M$ & $\lambda$ \\
\hline 
GM1\cite{GM} & 11.79         & 7.149         & 4.411               & 0.005894       & -0.006426 \\
GM3\cite{GM} & 9.927         & 4.820         & 4.791               & 0.017318       & -0.014526 \\
\hline
\end{tabular}
\end{center}
\caption{Set of parameters for the NLW model.}
\end{table}

\begin{table}
\begin{center}
\begin{tabular}{llcccccc}
\hline
        &                        &  $M_{\max}$ &  $M_{b \max}$ &  $R$ & ${\cal E}_0$ & ${\cal E}_{\min}$  & ${\cal E}_{\max}$ \\
Star type & Model                 & ($M_\odot$) & ($M_\odot$)   & (Km) & $(fm^{-4})$  &  $(fm^{-4})$       &  $(fm^{-4})$    \\
\hline
Hadron   & GM1n                   &  2.390       &  2.892        & 11.992 & 5.595 & - & - \\
Hadron   & GM1nh                  &  2.006       &  2.325        & 11.851 & 5.908 & - & - \\
\hline
Hybrid withou m.p. & GM1n$\times$SU(2)set1 &  1.835       &  2.108        & 11.259 & 6.464 & - & 1.241 \\
Hybrid withou m.p. & GM1n$\times$SU(2)set2 &  2.227       &  2.638        & 13.085 & 4.810 & - & 5.689 \\
Hybrid withou m.p. & GM1nh$\times$SU(3)RKH  &  1.970       &  2.276        & 12.542 & 6.615 & - & 6.607 \\
Hybrid withou m.p. & GM1nh$\times$SU(3)HK   &  1.906       &  2.189        & 12.821 & 4.538 & - & 4.000 \\
\hline
Hybrid with m.p. & GM1nh$\times$SU(3)RKH  &  1.945       &  2.242        & 12.568 & 4.979 & 3.454 & 7.797 \\
Hybrid with m.p. & GM1nh$\times$SU(3)HK   &  1.909       &  2.192        & 12.666 & 4.876 & 2.357 & 5.023 \\
\hline
Hadron   & GM3n                   &  2.042       &  2.421        & 10.933 & 7.048 & - & - \\
Hadron   & GM3nh                  &  1.710       &  1.946        & 10.980 & 7.151 & - & - \\
\hline
Hybrid without m.p. & GM3n$\times$SU(2)set1 &  1.836       &  2.110        & 11.287 & 6.464 & - & 1.303 \\
Hybrid without m.p. & GM3n$\times$SU(2)set2 &  2.018       &  2.381        & 11.484 & 8.300 & - & 8.295 \\
Hybrid without m.p. & GM3nh$\times$SU(3)RKH  &  1.710       &  1.946        & 10.977 & 7.161 & - & 11.895 \\
Hybrid without m.p. & GM3nh$\times$SU(3)HK   &  1.710       &  1.946        & 10.972 & 7.179 & - & 11.119 \\
\hline
Hybrid with m.p. & GM3nh$\times$SU(3)RKH  &  1.704      &  1.938        & 11.176 & 6.820 & 5.424 & 13.437 \\
Hybrid with m.p. & GM3nh$\times$SU(3)HK   &  1.700      &  1.934        & 11.198 & 6.772 & 4.772 & 12.797 \\
\hline
\end{tabular}
\end{center}
\caption{Maximum gravitational mass $M_{max}$, baryonic mass $M_{b max}$, and radius $R$.
$\varepsilon_0$ is the central energy density,
$\varepsilon_{\min}$ is the energy density where the hadron phase ends, and
$\varepsilon_{\max}$ is the energy density where the quark phase begins.}
\end{table}



\begin{thebibliography}{}
%
\bibitem{prakash} M. Prakash, I. Bombaci, P.J. Ellis, J.M.
Lattimer and R. Knorren, Phys. Rep. {\bf 280}, 1 (1997).

\bibitem{glen} N.K. Glendenning, Compact Stars, Springer-Verlag, New-York,
2000.

\bibitem{tov} R.C. Tolman, Phys. Rev. {\bf 55} (1939) 364; J.R. Oppenheimer and
G.M. Volkoff, Phys. Rev. {\bf 55} (1939) 374.

\bibitem{hadronicas} A.L. Esp\'{\i}ndola and D.P. Menezes, Phys. Rev. C {\bf 65}, 
045803 (2002);
A.M.S. Santos and D.P. Menezes, Phys. Rev. C {\bf 69}, 045803 (2004); R. Cavangnoli 
and D.P. Menezes, Braz. J. Phys. B {\bf 35}, 869 (2005).

\bibitem{quarkionicas} D.P. Menezes and D.B. Melrose, Publ. Astr. Soc. Aust. 
{\bf 22}, 292 (2005), D.P. Menezes, C. Provid\^encia and D.B. Melrose, 
J. Phys. G: Nucl. Part. Phys. {\bf 32}, 1981 (2006).

\bibitem{schaff} J. Schaffner-Bielich, J. Phys. G {\bf 31}, S651 (2005);
G. Pagliara and J. Schaffner-Bielich, Phys. Rev.  D {\bf 77} (2008) 063004.

\bibitem{lugones} G. Lugones and I. Bombaci, Phys. Rev. D {\bf 72} (2005) 
065021; G. Lugones et al. Phys. Rev D {\bf 80} (2009) 045017.

\bibitem{outros} K. Schertler et al., Phys. Rev. C {\bf 60} (1999) 025801;
M. Baldo et al., Phys. Lett. B {\bf 562} (2003) 153;
I.A. Shovkovy et al., Phys. Rev. D {\bf 67} (2003) 103004;
M. Buballa et al., Phys. Lett.  B {\bf 595} (2004) 36;
T. Kahn et al., Phys. Lett.  B {\bf 654} (2007) 170;
A. Steiner et al., Phys. Lett. B {\bf 486} (2000) 239;
F. Yang and H. Shen, Phys. Rev. C {\bf 77}, 025801 (2008).

\bibitem{hibridas} D.P. Menezes and C. Provid\^encia, Phys. Rev. C {\bf 68}, 
035804 (2003);
D.P. Menezes and C. Provid\^encia, Phys. Rev. C {\bf 70},058801 (2004);
D.P. Menezes and C. Provid\^encia, Phys. Rev. C {\bf 69}, 045801 (2004).

\bibitem{constanca} I. Bombaci, D. Logoteta, P. K. Panda, C. Provid\^encia 
and I. Vidana, Phys. Lett. B {\bf 680}, 448 (2009).

\bibitem{fraga} B.W. Mintz, E.S. Fraga, G. Pagliara and J. Schaffner-Bielich,
Phys. Rev. D {\bf 81}, 123012 (2010).

\bibitem{vos} T. Tatsumi, M. Yasuhira and D.N. Voskresensky, Nucl. Phys.
A {\bf 718} (2003) 359c.

\bibitem{maruyama} T. Maruyama et al. Phys. Rev. D {\bf 76} (2007) 123015

\bibitem{vos2} D.N. Voskresensky,  M. Yasuhira and T. Tatsumi, Phys. Lett. 
B {\bf 541}, 93 (2002), Nucl. Phys. A {\bf 723} 291 (2003).

\bibitem{bag} A. Chodos, R.L. Jaffe, K. Johnson, C.B. Thorne and V.F.
Weisskopf, Phys. Rev. D {\bf 9}, 3471 (1974).

\bibitem{sw} B. Serot and J.D. Walecka, {\em Advances in Nuclear
Physics} {\bf 16}, Plenum-Press, (1986) 1.

\bibitem{njl} Y. Nambu and G. Jona-Lasinio, Phys. Rev. {\bf 122}, 345 (1961);
{\bf 124}, 246 (1961).

\bibitem{yatsutake} N. Yasutake and K. Kashiwa, Phys. Rev. D {\bf 79}, 
043012 (2009).  

\bibitem{pasta_alpha} S.S. Avancini, C.C. Barros, D.P. Menezes and C. 
Provid\^encia, Phys. Rev. C {\bf 82}, 025808 (2010).

\bibitem{buballa} M. Buballa, Phys. Rep. {\bf 407} (2005) 205.

\bibitem{HK} T. Hatsuda and T. Kunihiro, Phys. Rep. 247 (1994) 221.

\bibitem{RKH} P. Rehberg, S.P. Klevansky, and J. H\"ufner, Phys. Rev. C {\bf 53} (1996) 410.

\bibitem{GM} N.K. Glendenning, S.A. Moszkowski, Phys. Rev. Lett. {\bf 67}, 2414 (1991).

\bibitem{bps} G. Baym, C. Pethick and P. Sutherland, Astrophys. J. ({\bf 170}), 
299 (1971).

\bibitem{schramm} S. Schramm and V.A. Dexheimer, to appear in Int. J. Mod. 
Phys. D.

\bibitem{vos3} Y.B. Ivanov, A.S. Khvorostukhin, E.E. Kolomeitsev,
V.V. Skokov, V.D. Toneev and D.N. Voskresensky, Phys. Rev. C {\bf 72}, 025804 
(2005).

\bibitem{cottam1} J. Cottam, F. Paerels and M. Mendez, Nature \textbf{420}, 51 (2002).

\bibitem{cottam2} J. Cottam, et al, arXiv:0709.4062v1 [astro-ph].

\bibitem{sanwal1} D. Sanwal, G.G. Pavlov, V.E. Zavlin, and M.A. Teter, 
Astrophys. J. Lett. \textbf{574}, L61 (2002).

\bibitem{bursts} D.P. Menezes, D.B. Melrose, C. Provid\^encia and K. Wu,
Phys. Rev. C {\bf 73}, 025806 (2006).

\bibitem{matt1} M. Alford, M. Braby, M. Paris and S. Reddy, arXiv:nucl-th/0411016v2.

\bibitem{matt2} M. Alford, D. Blaschke, A. Drago, T. Klahn, G. Pagliara and J. Schaffner-Bielich, arXiv:astro-ph/0606524v2.

\bibitem{matt3} O. Benhar and R. Rubino, A\&A {\bf 434}, 247-256 (2005).
%
\end{thebibliography}
\end{document}